\documentstyle[12pt]{article}
\begin{document}
\begin{flushright}
1996/1375
\end{flushright}  
\vspace{1 cm}
\begin{center}
\Large \bf The Equivalencce of Darmois-Israel- and Distributional-method    
for Thin Shells in General Relativity 
\end{center}
\vspace{1 cm}
\begin{center}
\Large R. Mansouri\dag \footnote{On Sabbatical leave from Department of 
Physics, Sharif University of Technology, P.O.Box 11365-9161, Tehran, Iran.}
, M. Khorrami\ddag     
\end{center}
\normalsize \dag Universit\"at Potsdam, Mathematisches Institut, Kosmologie 
Gruppe,\\ PF 60 15 53, 14415 Potsdam, Germany.\\  
\ddag University of Tehran, Department of Physics, Tehran, Iran.\\and\\
Institute for Studies in Physics and Mathematics, P.O.Box 5531, Tehran, Iran. 
\vspace{1 cm}
\begin{center}
\large {\bf  Abstract}
\end{center} 
A distributional method to solve the Einstein's field equations for thin
shells is formulated. The familiar field equations and jump conditions 
of Darmois-Israel formalism are derived. A carefull analysis of the 
Bianchi identities shows that, for cases under consideration, they make sense
as distributions and lead to jump conditions of Darmois-Israel formalism.    
\newpage 
\newcommand{\be}{\begin{equation}}
\newcommand{\ee}{\end{equation}}
\newcommand{\bea}{\begin{equationarray}}
\newcommand{\eea}{\end{equationarray}}
\newcommand{\bibi}{\bibitem}
\renewcommand{\baselinestretch}{1.3}
\newpage

\section{Introduction}
The study of hypersurfaces of discontinuity in general relativity has
begun in early twenties[1-3]. But it has been revived
through new questions raised in cosmology and black-hole physics.  
Domain walls separating two coexisting different phases in inflationary 
scenarios[4], bubble dynamics[5], wormholes[6], signature changes[7], 
and interior
structures of black-holes[8] are just some of the recent applications of
the thin shell formalism of general relativity. \\
The traditional and 
mostly used method of handling such problems is that of Darmois-Israel (DI),
 based on the Gauss-Kodazzi decomposition of space-time[9, 10]. It expresses
the surface properties in terms of the jump of extrinsic curvature
across the layer directly as functions of the layer's intrinsic 
coordinates. Thus the four-dimensional coordinates may be chosen freely
and independently, adapted to the symmetry requirements, on the two 
sides of the layer. This is the very practical advantage of that method 
which has found its final formulation in the outstanding paper of Israel[9]. 
The geometric
conditions for the layer to be considered as a boundary of two different       
manifolds glued together at this boundary are first formulated by
Darmois[3]. Those are minimum conditions which has been assumed by 
Israel. There are other conditions formulated by Lichnerowicz[11, 12], 
which means basically continuous coordinates across the layer, and seems
to be necessary for using distributional tensor calculus. It is interesting
to note that Sen[2], in this relatively unknown paper, uses the same 
conditions, without any further discusion,
derives the reduced Einstein's equations for the general case, and solve
it for the spheric symmetric 2+1 dimensional mass distribution.  
The O'Brien-
Synge conditions[13, 14] are in most cases equivalent to Lichnerowicz ones, so  
we are not going to consider them here[15]. Because of  
the restrictive choice of coordinates the Lichnerowicz conditions are
not usually used. But there are cases where the distributional method, 
 and therefore the Lichnerowicz conditions, are more suitable for
calculation. The case of cylindrically symmetric thin layer, for example, 
has been solved by distributional method using these conditions[16].\\ 
There have been attempts to formulate the problem of thin layer in 
general relativity using distributional methods, familiar in other
part of physics. In fact many    
physical systems, classical and relativistic, undergo very rapid 
transitions of their state of motion. Think of shock waves in hydrodynamics.
Although the state of the system need not be described by 
discontinuous functions of space and time, or by functions having 
discontinuity in their first or second derivatives, but a mathematical 
description of the system which is based on distribution valued states
of the system give an accurate picture of some important aspects of the 
physical problem. Usually such a description is more amenable to 
treatment than the treatment which contains a smooth description of 
the physical state. This has not only been used in classical hydrodynamics,
 but also bas been applied in the relativistic case. Lichnerowicz [11, 12] 
 has given a discussion of hydrodynamic and gravitational 
shock wave problems by using curvature tensors for space-time which contain
a Dirac $\delta$ function with support on a submanifold. Y.Choquet-Bruhat
has used similar methods to treat high-frequency gravitational waves
[17]. 
 
Rapid changes of physical quantities occur also in electromagnetism. One 
might have, for example, some charge distribution which is confined
to a one- or two-dimensional region of space small compared 
with characteristic distances of the problem. This distribution of charge
can be replaced by a concentrated source, and the problem can be 
formulated in the sense of distributions, and not of smooth functions.
There is a natural mathematical framework in electromagnetism. Recall that
linear operations, including differentiations, make sense when applied to 
distributions. Hence, Maxwell's equations, by virtue of their linearity 
in both fields and sources, make sense as equations on these distributions.
This means that the machinery of distribution theory is available in
electromagnetism, and guarantee that distributional Maxwell fields with 
distributional charge-currents make physical sense, and at the same 
time gives a well defined and detailed sense in which a distributional charge 
density must approximate our real smooth charge density distribution. 

In general relativity, where the field equations are non-linear, the use 
of distributional objects seem not to be trivial. Although Raju[18]
claims to give an analytical formalism to deal with the occurrence 
of jump discontinuities in the metric across a hypersurface, using 
a non-linear theory of distributions[18], his method does
not seem to be conclusive. The application he mentions is for a 
continuous metric where no non-linear distributional operations 
are needed. Anyhow, here the mathematical framework cannot be 
as simple as for electromagnetism. 
Efforts to implement distributional methods in general relativity goes 
back to works of Papapetrou and Treder[20]. Nariai[21], aiming to understand
the O'Brien-Synge junction conditions, and demanding the Einstein's tensor
to be free of $\delta$ functions, uses continuous metrics and so brings in 
distributions in the formalism, which is consequently used by Kumar[22]. 
Papapetrou
and Hamoui[23, 24] then try to formulate a general method with
application to spherical symmetric thin layer. Their method has then
been reformulated and corrected by Evans[25]. Lichnerowicz [12] gives a
detailed mathematical analysis of distributions in curved space-time and
comes to the conculsion that the classical properties of the covariant
derivatives and all of the corresponding formulas are valid for tensor
distributions. Barrabes [26] uses tensor distributions specially to include
null hyper-surfaces in the shell dynamics.\\
Taub[27] is 
interested in relativistic hydrodynamics and shock waves but also discusses 
the previous accounts on the concentrated 2- and 3-dimensional sources. 
Israel[28] and Taub[27] give a formulation for a 2-dimensional concentrated
mass distribution. These attempts have been criticized by Geroch
and Traschen[28] who give an extensive and thorough analysis of concentrated
mass distributions in general relativity. This work is a milestone in all the
discussions about the validity of distributional Einstein's field equations 
and its applications to concentrated sources. There the authors define
some regularity conditions for metrics, for which the  distributional
tensor calculations are allowed. Hence, for example, the line source
case should be handled with care, although some authors have criticized 
it[30]. The (2+1)-dimensional case, as a result of this work, should not 
arise any problem, as far as the continuity of the metric is assured. 
We have thereafter used distributional tensors to
solve the Einstein's equations directly, without any use of Gauss-Kodazzi
decompositions. Although the coordinates have to be prepared to make
them continuous at the hypersurface of discontinuity, but the method 
has been applied easily for several cases[16, 31, 32].   
Naturally, there should be no difference in the results using either of 
the distributional- or Gauss-Kodazzi-method. But the complete equivalence
has never been demonstrated explicitly, and the role of jump conditions  
by using the distributional method has never been clearly stated.\\ 
With the results of Geroch and Traschen in mind, we show here 
that all the dynamical- and constraint-equations derived by the DI-formalism 
results very naturally in the distributional method, without
any needs to define a new covariant derivative.\\ 
In section 2 we review shortly the D-I formalism and give
the necessary formulae. Section III begins first with the formulation 
of the distributional method. Covariant derivative of distributions 
and some useful formulae are  
given in section 3.1, and the Einstein's equations for a thin shell in
section 3.2. Section 3.3 deals with the conservation laws and the Bianchi 
identities. In this section we will see the full equivalence of the 
two methods. We end with a conclusion in section 4.\\  
{\bf Conventions and definitions}:\\  
We use the signature $(-+++)$, and follow the curvature conventions of
Misner, Thorn, and Wheeler (MTW)[10]. However, our sign convention for 
extrinsic curvature is that of Israel[9], which is the opposite of MTW. The
greek indices rum from $0$ to $3$ and latin indices from $1$ to $3$. A
semicolon indicates covariant derivatives with respect to either the four-
metric of the whole space-time or to the three-metric of the layer. There will,
 however, be no confusion because the kind of indices and objects 
used makes the 
difference transparent. The symbol $\nabla^\pm$ denotes the covariant 
derivative with respect to either of the metrics of partial manifolds  
$M^\pm$ which are to be glued together. \\   
The square brackets $[F]$ are used
to indicate the jump of any quantity $F$ at the layer, and 
bars $\overline F$ the arithmetic mean of it. As we are going to work 
with distributional valued tensors, there may be terms in a tensor quantity
$F$ proportional to some $\delta$-function. These terms are indicated
by $\breve{F}$. 

\section{Darmois-Isreal Formalism}
Assume two space-times $M^{+}$ and $M^{-}$ with boundaries $\Sigma^+$ and 
$\Sigma^-$.
 $M^+$ and $M^-$ may have been cut from space-times $M_1$ and $M_2$,
respectively, but this is
irrelevant for our task of glueing these together. Coordinates on the two
space-time manifolds are defined independently as $x^{\mu}_+$ and $x^{\mu}_-$,
 and the metrics denoted by $g^{+}_{\;\alpha \beta}(x^{\mu}_+)$ and
$g^{-}_{\;\alpha \beta}(x^{\mu}_-)$. The induced metrics on the boundaries
are called $g^{+}_{\;ij}(\xi^{k}_+)$ and $g^{-}_{\;ij}(\xi^{k}_-)$,
 where $\xi^{k}_{\pm}$ 
 are intrinsic coordinates on $\Sigma_{\pm}$, respectively. Bringing    
these 3- and 4-dimensional quantities in connection is trivially done
with the help of tetrads defining on $\Sigma$[34].\\ 
Now, to paste the manifolds together we demand the boundaries to
be isometric having the same coordinates $$\xi^{k}_+ = \xi^{k}_- = \xi^{k}.$$
The identification $$\Sigma_+ = \Sigma_- =: \Sigma$$ gives us the single 
glued manifold $M = M_+ \bigcup M_-$.\\ 
This is the minimum requirement for glueing two manifolds together. 
Formulated as    
\be     [g_{ij}] = 0                                  \ee
gives together with the continuity of the second fundamental form on $\Sigma$  
\be      [K_{ij}] = 0                                   \ee 
the Darmois conditions. Both conditions should be satisfied if $\Sigma$
 is just a boundary surface. But in case of a thin shell we do not expect 
the second condition to be satisfied. In fact, the matter content of 
the shell should lead to a jump in the extrinsic curvature $K_{ij}$. \\
The condition (1) leaves the coordinates in $M^\pm$ free. If we assume
the continuity of the coordinates $x^\mu_{\pm}$ at $\Sigma$ we then
have to require
\be          [g_{\mu \nu}] = 0,                         \ee  
which together with the corresponding equation for derivatives of the metric 
\be         [\frac{\partial g_{\mu \nu}}{\partial x^\alpha}] = 0                         \ee
gives the Lichnerowicz conditions. In the following we just assume 
the condition (1) or (3) respectively.\\ 
 On $\Sigma$ we define a three-bein
$$e_i = \frac{\partial}{\partial \xi^i}$$ having the components
\be
e^{\mu}_{i} = \frac{\partial x^{\mu}}{\partial \xi^{i}} 
\ee
The induced metric on $\Sigma$ is given by the scalar product
\be
g_{ij} = e_i \cdot e_j = g_{\mu \nu} e^{\mu}_{i}e^{\nu}_{j}
\ee
Note that, because of the assumed isometry, this metric is the same on 
both faces $\Sigma_{+}$ and $\Sigma_{-}$. We note that the subscripts of the
three-beins on $\Sigma$ are not the component indices, but as this distinction
is trivial we prefer for the sake of simplicity not to use parentheses to
distinguish them, as is usually done.\\ 
We choose the parametric equation for $\Sigma$ in the form
\be
\Phi(x^\mu(\xi^i)) = 0,  
\ee 
having the unit normal four-vector $n^\mu$ given by
\be
n_\mu = \alpha ^{-1} \partial_\mu \Phi,                              \ee
where
\be
\alpha = \pm 
\sqrt{(|g^{\nu \gamma} \frac{\partial f}{\partial x^\nu}
\frac{\partial f}{\partial x^\gamma}|)}. 
\ee
Therefore      
\be              n_\mu e^ \mu _i = 0,            \ee
and 
\be  n_{\mu} n^{\mu} =  \epsilon,    \ee  
where $\epsilon = + 1$ or $-1$ for $\Sigma$ to be time- or space-like,
respectively. We suppose $n^\mu$ directed from $M^-$ to $M^+$, i.e.   
in the direction of increasing a space- or time-like coordinate corresponding
to time- or space-like $\Sigma$. Therefore we have to take the 
positive (negative) sign in (9) for time- (space-)like $\Sigma$. This choice
gives us the useful relation $$\mbox{sign}\;\alpha= \frac{| \alpha|}{
\alpha}=\epsilon.$$  
The choise of Lichnerowicz condition (3) makes it possible to have a unique 
 normal vector for each case. As we want to concentrate on the formulation of 
the distributional method and avoid any undue complications, we leave aside 
the case of null hypersurfaces.\\  
Now, in general the metrics in $M^+$ and $M^-$ need not to be continuous at 
$\Sigma$, but they could be. However, the normal extrinsic curvature (second
fundamental form) is not continuous for a thin shell. It is defined by
\begin{eqnarray} 
K^\pm_{\;ij} &=& e^{\mu}_{\;i}e^\nu_{\;j} \nabla^\pm_{\; \mu} n_{\nu} =  
  - n_{\mu}e^\nu_{\; j} \nabla^\pm_{\; \nu} e^\mu_{\;i} \nonumber \\ 
  &=& -n_{\mu}e^\nu_{\; i} \nabla^\pm _{\;\nu} e^{\mu}_{\;j} = K^\pm_{\;ji} 
\end{eqnarray}
Now, we have all the prerequisites to write the Einstein's equation for
the hypersurface. These are 10 equations which will be written in
components normal and tangent to the hypersurface. The first and second
contracted Gauss-Kodazzi equations are [9, 10]  
\be
G_{\mu \nu}n^{\mu}n^{\nu} = \frac{1}{2}(K^2 -K_{ij}K^{ij}-\epsilon\; ^3 
\negthinspace R)
\ee
\be
G_{\mu \nu}e^{\mu}_{\;i} n^{\nu} = K^{j}_{\;i;j}- K_{,i},     \ee
where $^3 \negthinspace R$ and $^3 \negthinspace G$ are the Ricci scalar 
and Einstein tensor of the
three metric $g_{ij}$, respectively. 
Now, to discover the effect the energy-momentum tensor $S_{ij}$ of 
$\Sigma$ on the space-time geometry, we perform a "pill-box" integration
of Einstein's equations across $\Sigma$:
\be
S_{\mu \nu} = \lim_{\Sigma \rightarrow 0}\int^{\Sigma}
_{-\Sigma} \left(T_{\mu \nu} - g_{\mu \nu}
\frac{\Lambda}{\kappa} \right)\,dn
 =  \frac{1}{\kappa} \lim_{\Sigma \rightarrow 0} 
\int^{\Sigma}_{-\Sigma} G_{\mu \nu} \,dn ,
\ee
where n is the proper distance through $\Sigma$ in the direction of the normal 
$n_\mu$.\\ 
$S_{\mu \nu}$ is the associated 4-tensor of energy momentum of
the shell. The equations (13, 14) have the physical meaning that no moment 
associated with the surface layer flows out of $\Sigma$. 
 Therefore $S_{\mu \nu}$ vanishes off the hypersurface $\Sigma$, which
is expressed as 
\be
S_{\mu \nu} n^\nu = 0                                      \ee
 
The energy momentum 4- and 3-tensors are related as   
\be
S^{\mu \nu} = e^\mu_i e^\nu_j S^{ij}                             \ee
The covariant derivative of such a tensor relative to the corresponding
connections is given by [9]  
\be
\nabla^\pm_{\;\nu} S^{\mu \nu} = e^\mu_i S^{ij}_{\; ;j} - \epsilon S^{ij} 
K^\pm_{\;ij} n^{\mu},   
\ee
which leads to the following useful relation
\be
e^\mu_{\; i}\nabla _\nu S_\mu^{ \nu} = S^j_{\; i:j} 
\ee  
Similarly we can associate to the 3-tensor $K_{i j}$  defined on $\Sigma$,
 the corresponding 4-dimensional tensor:
\be 
K^{\mu \nu} = K^{i j} e^\mu_i e^\nu_j,                          \ee 
satisfying  
\be 
K^{\mu \nu} n_\nu = 0.                                      \ee 
The remaining components of the Einstein's equations lead to the
following non-vanishing result 
\newpage
$$
\lim_{\Sigma \rightarrow 0} \int^\Sigma_{-\Sigma} G_{\mu \nu}
e^{\mu}_ie^{\nu}_j \;dn = 
$$
\be    
\epsilon \left( [K_{ij}] - g_{ij}[K] \right)= \kappa S_{ij}  
\ee     
This distributional equivalent of Einstein's equations is called {\it Lanczos} 
equation, which partly determines the dynamic of the thin shell. The
other dynamical equations come from the defining equation of matter contents
of the shell. Now, the two Gauss-Kodazzi equations act as constraints. The 
first one (13) is the so-called "Hamiltonian "- and the second one (14) 
the "ADM"- constraint. 
Note however that these 
equations are valid in $M^+$ and $M^-$ on taking the limits
as one approaches the layer $\Sigma$. Therefore we are actually faced
with 8 equations, the sum and difference of which give us  the 
junction conditions. 
The Hamiltonian constraint along
with the Einstein's or Lanczos equations then give the 
{\it evolution identity}: 
\be
S^{ij} \overline K_{ij} = - \left[T_{\mu \nu}n^{\mu}n^{\nu} - 
\Lambda/\kappa \right]                              \ee 
and 
\be
^3\negthinspace R + (\overline K_{ij} \overline K^{ij} - K^2) 
= 2\epsilon\kappa \overline {(T_{\mu \nu}
n^{\mu}n^{\nu} - \Lambda /\kappa)}+   
\frac{\epsilon \kappa^2}{4} (S^{ij}S_{ij} - \frac{S^2}{2}).  
\ee   
The ADM constraint gives the {\it conservation identity}
\be
S^i_{\;j;i} = -\epsilon [T_{\mu \nu}n^\mu e^\nu_j]                  \ee 
and
\be  
\overline K^i_{\; j;i} - \overline K_{,j} = \kappa 
\overline {(T_{\mu \nu} n^\mu e^\nu_i)}     \ee
Not all of these jump conditions are independent. Usually one  
takes the evolution identity (23) and the conservation identity (25)
as the proper junction conditions, which in addition to the Lanczos 
equation should be satisfied[35].

\section{Distributional Method}
Here we intend to give a formulation of the Einstein's
equations for the case where there exists a hypersurface of concentrated
source immersed in an otherwise arbitrary space-time, not necessarily
vacuum. We assume the metric to be continuous at the hypersurface:  
\be     [g_{\mu \nu}] = 0                        \ee
Otherwise we would have to consider non-linear operations of distributions
such as $\delta \theta$ or $\delta \delta$. The disadvantage of having a
continuous metric across the shell pays off by the simplicity of the 
method to calculate specific solutions[16, 31, 32].\\ 
Write the metric in the following form
\be
g_{\mu \nu} = g^+_{\;\mu \nu} \theta(\Phi(x)) + g^-_{\;\mu \nu}
 \theta(-\Phi(x))  
\ee  
where $\theta$ is the step function and  
\be
g^+_{\;\mu \nu}|_{\Phi(x) =0} = g^-_{\;\mu \nu}|_{\Phi(x)=0}     \ee  
This condition guarantee the smoothness of the metric on the hypersurface.
Should this not be the case we try a coordinate transformation $x= x(x')$
 having a jump in the first derivative:
\be
\frac{\partial x^\mu}{\partial x'^\rho} = \alpha ^{+ \mu}_{\; \rho} 
\theta(\Phi(x)) + 
\alpha^{- \mu}_{\; \rho} \theta(-\Phi(x))
\ee
The condition for the new metric to be continuous comes out to be 
\be
\alpha^{+ \mu}_{\; \rho} \alpha^{+ \nu}_{\; \sigma}\; g^+_
{\;\mu \nu}|_{\Phi(x)=0} = 
\alpha^{- \mu}_{\;\rho} \alpha^{- \nu}_{\;\sigma}\; g^-_
{\;\mu \nu}|_{\Phi(x)=0}  
\ee
We assume from now on that   
the metric is smooth everywhere, $C^1$ at the hypersurface and
$C^{\infty}$ on both sides of it.\\  
Although the metric is continuous on $\Sigma$, its derivatives, and 
so the corresponding connections, are discontinuous. Nevertheless 
the connection corresponding to the metric $g_{\mu \nu}$ can be written 
in the following compact form: 
\begin{eqnarray}  
\Gamma^\rho_{\; \mu \nu} &=& \frac{1}{2}g^{\rho \sigma}\left(
g_{\mu \sigma , \nu} + g_{\nu \sigma , \mu} - g_{\mu \nu , \sigma}
\right) \nonumber \\
   &=& \theta(\Phi(x)) \Gamma^{+\rho}_{\quad\mu \nu}+\theta(-\Phi(x)) \Gamma
^{-\rho}_{\quad\mu \nu},  
\end{eqnarray}
where $\Gamma^{\pm \rho}_{\quad\mu \nu}$ are the ordinary connections on
$M^{\pm}$. The above connection has jump discontinuities on $\Sigma$.\\ 
To write the field equations for the hypersurface we need the 
formulation of the energy-momentum tensor of the shell. Generally it 
can be written in the form
\be
{\breve{T}}_{\mu \nu} = C S_{\mu \nu}\delta (\Phi(x))                     \ee
where $C$ is a constant to be calculated. We integrate the above equation
in the direction of the normal to the hypersurface
\be
\int {\breve{T}}_{\mu \nu}\,dn = C S_{\mu \nu} \int \delta (\Phi(x)) \,dn = 
C S_{\mu \nu}|\frac{dn}{d\Phi}|                                  \ee
Therefore, using the definition (15), we obtain  
\be      C = |\frac{d\Phi}{dn}|                                \ee
and
\be
{\breve{T}}_{\mu \nu} = S_{\mu \nu} |\frac{d\Phi}{dn}|\delta (\Phi(x))
\ee
Note that in the literature one usually takes $C=1$, which is correct just for 
special cases. But in 
general the factor C is necessary (see also [33]). 
Now, the derivative of $\Phi$ in the normal direction can be written in
terms of unit normal vector $n^\mu$:
\be   
C=|\frac{d\Phi}{dn}| = |n^\mu \partial_\mu \Phi| = |\epsilon \alpha|
=|\alpha|,
\ee
where we have used (8-11). Therefore (33) will be written in the form
\begin{eqnarray}
{\breve{T}}_{\mu \nu} &=& S_{\mu \nu} |n^\sigma \partial_\sigma 
\Phi |\; \delta(\Phi(x)) \nonumber \\
&=& CS_{\mu \nu}\delta(\Phi (x)) = |\alpha| S_{\mu \nu} 
\delta(\Phi(x))  
\end{eqnarray}   

\subsection{Covariant Derivative of Distributional valued Tensors}
There is no need to change of the ordinary concept of covariant
derivative, as it has been carefully shown by Lichnerowicz [12]. In fact, all
the known properties of covariant derivative and the corresponding formulae
in a pseudo-Riemannian manifold are valid for tensor distributions. 
But for the sake of convenience of calculation we refer
to some useful formulae.   
Consider first an arbitrary vector $A^{\mu}$ defined as
\be
A^\mu = \theta(\Phi) A^+ + \theta(-\Phi) A^-,                    \ee
where $A^\pm$ has the support on $M^\pm$.  
It is therefore useful to define the operator  
\be
\nabla = \theta(\Phi) {\nabla}^{+} + \theta (-\Phi) {\nabla}^{-}          
\ee 
We can know write the covariant derivative of a distributional valued vector
$A^\mu$ in terms of the covariant derivatives of its defining parts in
$M^\pm$. The following relation is easily obtained: 
\be 
A^\mu_{\; ;\nu} = \nabla_\nu A^\mu + [A^\mu]\;\partial_\nu \Phi \; \delta(\Phi)
\ee  
This relation can be generalized easily for a distributional tensor
of any rank. The covariant derivative of the tensor  
\be  
T^{(\rho)} = \theta (\Phi) T^{+(\rho)} + \theta (-\Phi)T^{-(\rho)},
\ee 
where $(\rho)$ stands for any number of indices, is 
calculated to be  
\be
T^{(\rho)}_{\quad ;\nu}=\nabla_\nu T^{(\rho)}+[T^{(\rho)}]\;\partial_\nu \Phi 
\;\delta (\Phi)
\ee 
In the case a tensor has the support on $\Sigma$ its covariant derivative
is in the usual form. Take the tensor $\breve{T}^{\mu \nu}$ from (33). Its
covariant derivative can be written
\begin{eqnarray} 
(\breve{T}^{\mu \nu})_{;\rho} &=& (CS^{\mu \nu})_{; \rho}
\; \delta(\Phi) + CS^{\mu \nu} \; (\delta(\Phi))_{,\rho} \nonumber \\
&=& (CS^{\mu \nu})_{; \rho}\; \delta(\Phi) + CS^{\mu \nu} \;
\partial_\rho \Phi \; \delta'(\Phi)
\end{eqnarray} 
We will need this relation later to discuss the conservation laws. 

\subsection{ The Field Equations} 
The Einstein's field equations are valid on both sides of the hypersurface as
usual. So we concentrate our procedure on $\Sigma$, 
where we expect the curvature and Einstein tensor to  
be proportional to $\delta$. That means in calculating the connection 
coefficients and the components of the Ricci tensor we can ignore terms not 
proportional to $\delta$. Hence, e.g., the terms in the Ricci tensor  
\be   
R_{\mu\nu} = \Gamma^\rho_{\mu \rho,\nu} - \Gamma^\rho_{\mu \nu,\rho} +
\Gamma^\sigma_{\mu \rho} \Gamma^\rho_{\sigma \nu} - \Gamma^\sigma_{\mu \nu}
\Gamma^\rho_{\rho \sigma} 
\ee  
proportional to $\Gamma$'s can be ignored. The only relevant terms are
\be
\breve{R}_{\mu \nu} = \breve{\Gamma}^\rho_{\mu \rho,\nu} - \breve{\Gamma}^\rho
_{\mu \nu,\rho}
\ee
Now, 
\be
\Gamma^\rho_{\mu \rho} = \frac{1}{2g} g_{,\mu} ,                          \ee
where $g$ is the determinant of the metric. The $\delta$ distribution 
can only occur in the second derivatives of the metric. Therefore
\be
\breve{\Gamma}^\rho_{\mu \rho,\nu} = \frac{1}{2g}\breve{g}_{,\mu \nu} 
\ee
Similarly, for the second term in the Ricci tensor we have
\be
\breve{\Gamma}^\rho_{\mu \nu,\rho} = \frac{1}{2}g^{\rho \sigma}  
\left(\breve{g}_{\sigma \mu, \nu \rho} + \breve{g}_{\sigma \nu,\mu \rho} -    
\breve{g}_{\mu \nu, \sigma \rho} \right) 
\ee
Having the metric in the form (28) we obtain
\be
\breve{g}_{\alpha \beta,\mu \nu} = [g_{\alpha \beta, \mu}](\partial_{\nu} \Phi )
\delta  (\Phi(x)) 
\ee
and 
\be
\breve{g}_{,\mu \nu} = [g_{,\mu}] \partial_{\nu} \Phi \delta(\Phi(x))  
\ee 
As the result we obtain for terms in the Ricci tensor proportional to $\delta$  
\begin{eqnarray}    
\breve{R}_{\mu \nu} &=& \left(\frac{1}{2g}g_{,\mu}] \partial_{\nu}\Phi 
-g^{\rho \sigma}\left( [g_{\sigma \mu,\nu}] + [g_{\sigma
\nu,\mu}] - [g_{\mu \nu, \sigma}] \right) \partial_{\rho} f\right)  
\delta(f(x)) \nonumber\\
&=& \left(\frac{1}{2g} [g_{,\mu}] \partial_{\nu}\Phi - [\Gamma^{\rho}_{\mu \nu}]
\partial_\rho \Phi \right) \delta(\Phi(x)) 
\end{eqnarray}    
This enable us to write the Einstein's equations for the layer:
\be
\breve{G}_{\mu \nu} = \kappa \breve{T}_{\mu \nu}                 
\ee 
Defining
\begin{eqnarray} 
Q_{\mu \nu} &=& (\alpha)^{-1}\left(\frac{1}{2g}[g_{,\mu}]\delta^{\rho}_{\nu} - 
[\Gamma^\rho_{\mu \nu}]\right)\partial_\rho \Phi \nonumber \\
&=& \left(\frac{1}{2g} [g_{,\mu}]\delta^{\rho}_{\nu}-
[\Gamma^\rho_{\mu \nu}]\right) n_\rho
\end{eqnarray}             
we obtain, using (38) and (52) for the energy momentum tensor 
the field equations in the 4-dimensional form  
\be
Q_{\mu \nu} - \frac{1}{2}g_{\mu \nu} Q =\epsilon \kappa S_{\mu \nu} , 
\ee
where $Q = Q_{\mu \nu} g^{\mu \nu}$, and we have used the relation 
$\epsilon=\frac{|\alpha|}{\alpha}$. Note that $Q_{\mu \nu}$ is a tensor 
with support on  $\Sigma$. This equation, for the time-like case, has 
been first derived, without using the hitherto unknown distributional
calculus, by Sen[2]. We would like, therefore, to coin it 
by {\it Sen equation}. 
The three dimensional form of the Sen equation is readily
obtained by decomposing it to tangential- and normal-components
to $\Sigma$. Multiplying (54) with $n^\mu$ we obtain        
\be
S_{\mu \nu}n^\nu = 0  ,                                   \ee 
which is the same relation as (16). 
This tell us immediately that the components
corresponding to $S_{\mu \nu}n^\mu n^\nu$ and $S_{\mu \nu}n^\mu e^\nu_i$
identically vanishes. To obtain the proper 3-dimensional components we 
notice first that
\be
Q_{ij}=Q_{\mu \nu} e^\mu_i e^\nu_j = - [\Gamma^\rho_{\mu \nu}] n_\rho
e^\mu_i e^\nu_j = [K_{ij}]. 
\ee 
Threfore, we obtain from the Sen equation
\be
Q_{i j} = \epsilon \kappa \left(S_{i j} - \frac{1}{2}g_{ij} S \right), 
\ee
which is equivalent to the Lanczos equation (22).\\
We have therefore seen that the explicit method of writing the Einstein's
field equations for a regular metric which is continuous without having
continuous derivatives leads to the equation (55) and is equivalent to  
the DI-formalism based on the Gauss-Codazzi formalism. In practice,
one begins with known solutions of the Einstein's equations in $M^\pm$, and
after making sure the continuity of the metric on $\Sigma$, tries to solve
the equations (55). In the following we will show that the jump 
conditions of DI-formalism follows from the Bianchi identities corresponding
to the metric (28), and are therefore implicit in the equations we have 
used. 
  
\subsection{Conservation Laws}   
We have now all the prerequisites to evaluate the Bianchi identities and
the conservation of energy momentum tensor of our pasted space-time. The
energy momentum of the whole space-time, including the cosmological
terms $\Lambda^{\pm}$ is
\be
T^{\mu \nu} = \breve{T}^{\mu \nu} + (T^{+ \mu \nu} - \Lambda^{+}/\kappa\; 
g^{+ \mu \nu})\theta(\Phi)+(T^{- \mu \nu} - \Lambda ^-/\kappa\; g^{- \mu \nu})  
\theta(-\Phi),
\ee
where $\breve{T}^{\mu \nu}$ is defined in (38). Having in mind that
the covariant divergences of $T^{\pm \mu \nu}$ and $g^{\pm \mu \nu}$
with respect to the corresponding connections vanishes, we obtain
\be
T^{\mu \nu}_{\quad ;\nu} = (\breve{T}^{\mu \nu})_{; \nu} + [T^{\mu \nu}- 
\Lambda/\kappa]\; \partial_\nu \Phi \; \delta(\Phi),    
\ee 
where we have used the relation (43). Now inserting for $\breve{T}^{\mu \nu}$
from (38) and using (44) we obtain an equation having terms proportional
to $\delta(\Phi)$ and $\delta'(\Phi)$. Each term vanishes independently. The 
term proportional to $\delta'(\Phi)$ gives
\be
(S^{\mu \nu} \partial_\nu \Phi) (n^\sigma \partial_\sigma \Phi) = 0,    \ee
which leads to the (56) and ensures the orthogonality 
of the energy-momentum tensor of 
$\Sigma$ to the hypersurface normal $n^\mu$. We use in the following this
relation to simplify the remaining calculations. The term proportional
to $\delta$ is  
\be
\left(CS^{\mu \nu})_{,\nu}  
+ CS^{\mu \rho} \Gamma^\nu_{\nu \rho} + 
CS^{\rho \nu} \Gamma^\mu_{\rho \nu} +  
+[T^{\mu \nu} - \Lambda/\kappa\; g^{\mu \nu}] 
\partial_\nu \Phi \right)\delta(\Phi)=0.  
\ee
We have left $\delta(\Phi)$ as proportionality factor to stress its 
influence specially on terms containing $\Gamma$'s and $g^{\mu \nu}$'s. 
Note that the third term containing $\Gamma$'s 
contains terms like $\theta \cdot \delta$, i. e. product 
of distributions. This is in analogy to the elementary problem of 
evaluating the electrostatic force on a sheet of charge[5].  
There the linearity of electrostatic equation resolve the ambiguity. But
how about our case where the Einstein equations are non-linear? We have 
already shown that in the case of thin shells the only terms contributing to
the Einstein tensor are the derivetives of the connection, or the second 
derivative of the metric, which appear linearly. It is then easily seen 
that is case of a layer the Einstein's equations leads to a Poisson-like 
equation corresponding to the Sen equation (55) for concentrated distribution,
 where the second derivative can be replaced by (50) and (51). 
Therefore, in analogy to the electromagnetic
case we can use the linearity of (55) to show that 
\be
\Gamma^\mu_{\rho \nu}\delta(\Phi) = \frac{1}{2}(\Gamma^{+\mu}_{\quad \rho \nu}  
+ \Gamma^{- \mu}_{\quad \rho \nu}) \delta(\Phi)
\ee
Using this result and multiplying the 
equation (62) with $n_\mu$, we obtain the component in the normal direction.
\be
(CS^{\mu \nu})_{,\nu} n_\mu + 
CS^{\rho \nu} \overline \Gamma^\mu_{\; \rho \nu} n_\mu
= \epsilon [T^{\mu \nu} - \Lambda/\kappa\; g^{\mu \nu}]n_\mu \partial_\nu \Phi.
\ee
Using the relations (63) and the definition of the extrinsic curvature (12), we
obtain the final result  
\be
S^{ij} \overline K_{ij} = \epsilon [T^{\mu \nu}n_\mu n_\nu - \Lambda/\kappa] 
\ee  
This is the evolution identity (23) derived as one of the
jump conditions in the Darmois-Israel method. Here it is just a
consequence of the Bianchi identities.
To obtain the remaining three equation we multiply (64) with $e^i_\mu$. 
Using (19) we obtain   
\be
S^i_{\;j;i} = -\epsilon\left[T_{\mu \nu} n^\mu e^\nu_i \right],  
\ee 
which is the conservation identity (25). It gives  
the conservation law for the energy momentum tensor of the layer.
We therefore see that our explicit distributional method of solving
the Einstein's equations gives all the dynamical and constraint equations
of Darmois-Israel method, and is therefore equivalent to it. 
\section{Conclusion}
We have seen that, based on the Lichnerowicz condition (3), a distributional
method can be formulated to solve the Einstein's field equations for a 
thin shell, which is equivalent to the Darmois-Israel formalism, 
and gives all the 
necessary equations and jump conditions formulated there. 
In fact, it has been shown that the jump conditions are consequence of 
the Bianchi identities, and therefore implicit in the formalism,   
once the Lichnerowicz condition is satisfied.   
This makes the distributional formalism  
easy to apply, specially when explicit solutions are to be found,
 and it pays off the disadvantage of the continuity of the 
coordinates across the shell.

\end{document}